\documentclass[twocolumn]{aastex701}

\usepackage{amsmath}
\usepackage{natbib}
\usepackage{graphicx}

\shorttitle{A Candidate Low-Mass Disk-Eclipsing Binary in UPK~13}
\shortauthors{Lin et al.}

\begin{document}

\title{A Candidate Low-Mass Disk-Eclipsing Binary in the $\sim$316~Myr Open Cluster UPK~13}

\author[0009-0008-9942-620X]{Jiamao Lin}
\email{linjm66@mail2.sysu.edu.cn}
\affiliation{School of Physics and Astronomy, Sun Yat-sen University,
Zhuhai 519082, China}
\affiliation{CSST Science Center for the Guangdong-Hongkong-Macau
Greater Bay Area, Sun Yat-sen University, Zhuhai 519082, China}

\correspondingauthor{Jiamao Lin}

\author[0000-0002-3935-2666]{Yongkang Sun}
\email{sunyk@ucas.ac.cn}
\affiliation{Department of Astronomy, University of Chinese Academy
of Sciences, Beijing 100049, People's Republic of China}

\author{Chengyuan Li}
\email{lichengy5@mail.sysu.edu.cn}
\affiliation{School of Physics and Astronomy, Sun Yat-sen University,
Zhuhai 519082, China}
\affiliation{CSST Science Center for the Guangdong-Hongkong-Macau
Greater Bay Area, Sun Yat-sen University, Zhuhai 519082, China}

\begin{abstract}  
UPK~13-c2 is a candidate member of the $\sim$316~Myr open cluster  
UPK~13 and was previously classified as a white dwarf + main-sequence  
(WD$+$MS) binary with 99.44\% confidence. We present multi-band  
photometric evidence that it is instead more plausibly a late-K/early-M binary  
with a misaligned circumbinary disk. The photometry reveals a  
flat-bottomed eclipse at $P=36.71$~days with an approximately  
achromatic $\sim$40\% flux decrement from the optical through $W1$, a  
reduced $W2$ depth, and a prominent mid-infrared excess at $W3/W4$.  
Two independent diagnostics strongly disfavor the WD$+$MS  
interpretation. First, the flat eclipse floor and 2.5-day ingress  
require complete occultation of an extended stellar component; a white  
dwarf would cross the disk edge in $\lesssim$2~hr and cannot naturally  
reproduce the observed multi-day trapezoid. Second, the difference  
spectrum is well fit by a single $\sim$4000~K thermal spectral energy  
distribution, favoring a late-K/early-M dwarf over a white dwarf as the  
occulted source. A decoupled SED decomposition yields a template-based  
late-K/early-M binary estimate of M1V+K9V with  
$M_{\rm tot}\approx1.4\,M_\odot$ and an overall systematic uncertainty  
of about 20\%. Under a sharp-edge eclipse model, forward modeling of  
the light curve favors an eccentric, spatially localized occulting  
structure. The light-curve morphology places UPK~13-c2 in the same  
geometric class as KH~15D and Bernhard-2. If cluster membership is  
confirmed, UPK~13-c2 may be the oldest known main-sequence  
disk-eclipsing binary.  
\end{abstract}


%% ======================================================================
\section{Introduction} \label{sec:intro}

In the rare class of disk-eclipsing binaries, periodic dimming arises
not from mutual stellar eclipses but from occultation by a misaligned
circumbinary disk. The prototype is KH~15D (V582~Mon)
\citep{Hamilton2001, Herbst2002, Winn2004, Winn2006, Chiang2004,
Poon2021}, a pre-main-sequence binary in NGC~2264 ($\sim$3~Myr) with
a precessing disk inclined $\sim$5--15$^\circ$ to the binary orbit.
\citet{Zhu2022} subsequently identified two additional KH~15D-like
systems from ZTF
(Bernhard-1, ZTF~J202055.22$+$381323.1;
Bernhard-2, ZTF~J071445.39$-$090152.1), and \citet{Hu2024, Hu2026}
spectroscopically confirmed Bernhard-2 as an eccentric MS$+$MS binary
($e=0.69$), establishing that flat-bottomed eclipses arise from the
\emph{complete} occultation of one binary component by the disk
edge. Additional disk-eclipsing systems have been catalogued among
T~Tauri and Herbig~Ae/Be stars
\citep{Bouvier2007, Rodriguez2016}, but all confirmed examples are
young ($\lesssim$20~Myr).

Theoretically, misaligned and even polar circumbinary
configurations are an expected outcome of binary--disk evolution,
particularly around eccentric binaries, where dissipative
realignment competes with secular precession driven by the
non-axisymmetric binary potential
\citep{Aly2015, Lai2023}. Such
geometries are a natural consequence of binary--disk coupling rather
than a pathological configuration. From an observational standpoint,
disk-eclipsing systems are particularly valuable: they provide
unusually direct constraints on the disk inner-edge structure,
on the magnitude and timescale of nodal precession, and on the
secular response of the disk to binary tidal torques, all
encoded in the long-term evolution of the eclipse profile
\citep{Winn2004, Winn2006, Chiang2004, Poon2021, Hu2026}. Yet the
known sample remains confined to very young systems
(KH~15D, $\sim$3~Myr; Bernhard-1/2, $\lesssim$20~Myr;
\citealt{Hamilton2001, Zhu2022, Hu2024, Hu2026}), leaving a substantial
age gap: it is not yet known whether comparable occulting
circumbinary structures can persist into the mature
main-sequence phase, where any surviving material would either
require unusually efficient preservation of primordial gas and dust
or signal a fundamentally different (e.g., second-generation)
origin.

While primordial circumstellar disks typically disperse within
$\lesssim 10$--$20$~Myr, rare ``Peter Pan'' disks around low-mass stars
can persist up to $\sim$50~Myr \citep{Silverberg2020}. The detection
of a dense circumbinary disk candidate around a $\sim$300~Myr
main-sequence binary, if confirmed, would raise a fundamental open
question in disk evolution: whether such strongly occulting structures
represent exceptionally long-lived primordial reservoirs preserved by
binary tidal truncation \citep{Lai2023}, or massive second-generation
debris disks entering an extreme collisional phase.

In this paper, we present multi-band photometric evidence that
UPK~13-c2 (ZTF J184419.33$-$175333.81; 2MASS~J18441933$-$1753336), classified by
\citet{Grondin2024} as a WD$+$MS candidate with
$P_{\rm WD+MS}=99.44\%$ and a candidate member of the $\sim$316~Myr
cluster UPK~13 \citep{CantatGaudin2020}, is more plausibly a
late-K/early-M binary occulted by a misaligned circumbinary disk. We test
the WD$+$MS hypothesis through two independent observational
diagnostics: the eclipse morphology and the shape of the difference
spectrum. We find that both strongly disfavor the WD$+$MS
hypothesis, and we argue that an MS$+$MS$+$disk model
self-consistently reproduces the photometric, geometric, and SED
constraints. If confirmed as a member of UPK~13, UPK~13-c2 would
offer a rare opportunity to test whether comparable occulting
circumbinary structures can persist to substantially older ages than
in currently confirmed examples. The paper is organized as
follows: Section~\ref{sec:data} describes the data;
Section~\ref{sec:analysis} presents the
analysis (light curve, SED, disk geometry, cluster kinematics);
Section~\ref{sec:discussion} discusses the WD$+$MS
classification, disk longevity, and cluster membership; and
Section~\ref{sec:summary} summarizes our findings and outlines
follow-up tests.

%% ======================================================================
\section{Observations and Data} \label{sec:data}

\subsection{Target and Multi-band Photometry} \label{subsec:target}

UPK~13-c2 (R.A.$=281\fdg0806$, Decl.$=-17\fdg8927$, J2000;
$G=17.62$~mag) is a Tier-2 WD$+$MS candidate from \citet{Grondin2024}. The key astrometric and photometric parameters
are summarized in Table~\ref{tab:params}. We assembled multi-band
photometry from ZTF DR23 \citep[$g$, $r$;][]{Bellm2019, Masci2019},
NEOWISE-R \citep[$W1$, $W2$;][]{Mainzer2014}, AllWISE \citep[$W3$,
$W4$;][]{Wright2010, Cutri2013}, 2MASS
\citep[$JHK_s$;][]{Skrutskie2006}, Pan-STARRS1
\citep[$izy$;][]{Chambers2016, Tonry2012}, and SPHEREx
spectrophotometry \citep[$0.75$--$5.0~\mu$m, $R\sim 40$--$130$;][]{Crill2020, Dore2018}. Single-epoch 2MASS ($\phi=0.90$) and
PS1 ($\phi=0.07$) measurements were verified to fall in the
out-of-eclipse phase window using the photometric ephemeris
established below.

We apply a Gaia DR3 parallax zero-point correction following
the magnitude-binned recipe of \citet{MaizApellaniz2022}, which is
tailored for high stellar-density regimes near the Galactic plane and
in clusters. For UPK~13-c2 ($G=17.59$, ecliptic latitude $5\fdg14$, 5-parameter
astrometric solution), the recipe yields a bias of $-38~\mu$as,
i.e.\ $\pi_{\rm corr}=1.051\pm0.133$~mas after adding a 10~$\mu$as
systematic floor in quadrature with the catalog uncertainty.  The
implied distance is $d=951^{+135}_{-105}$~pc, a 3.6\% reduction from
the uncorrected value.  As a cross-check, the \citet{Lindegren2021}
global recipe via the \texttt{gaiadr3\_zeropoint} package returns
$-33~\mu$as, agreeing with \citet{MaizApellaniz2022} at the
$\sim$0.005~mas level for this magnitude range.  Throughout the
remainder of the paper we adopt this value; all bolometric
luminosities, masses, semi-major axes, and cluster-membership
statistics reported below have been recomputed at $d=951$~pc.

\subsection{ZTF Optical Photometry} \label{subsec:ztf}

\begin{figure*}[t!]
\centering
\includegraphics[width=0.90\textwidth]{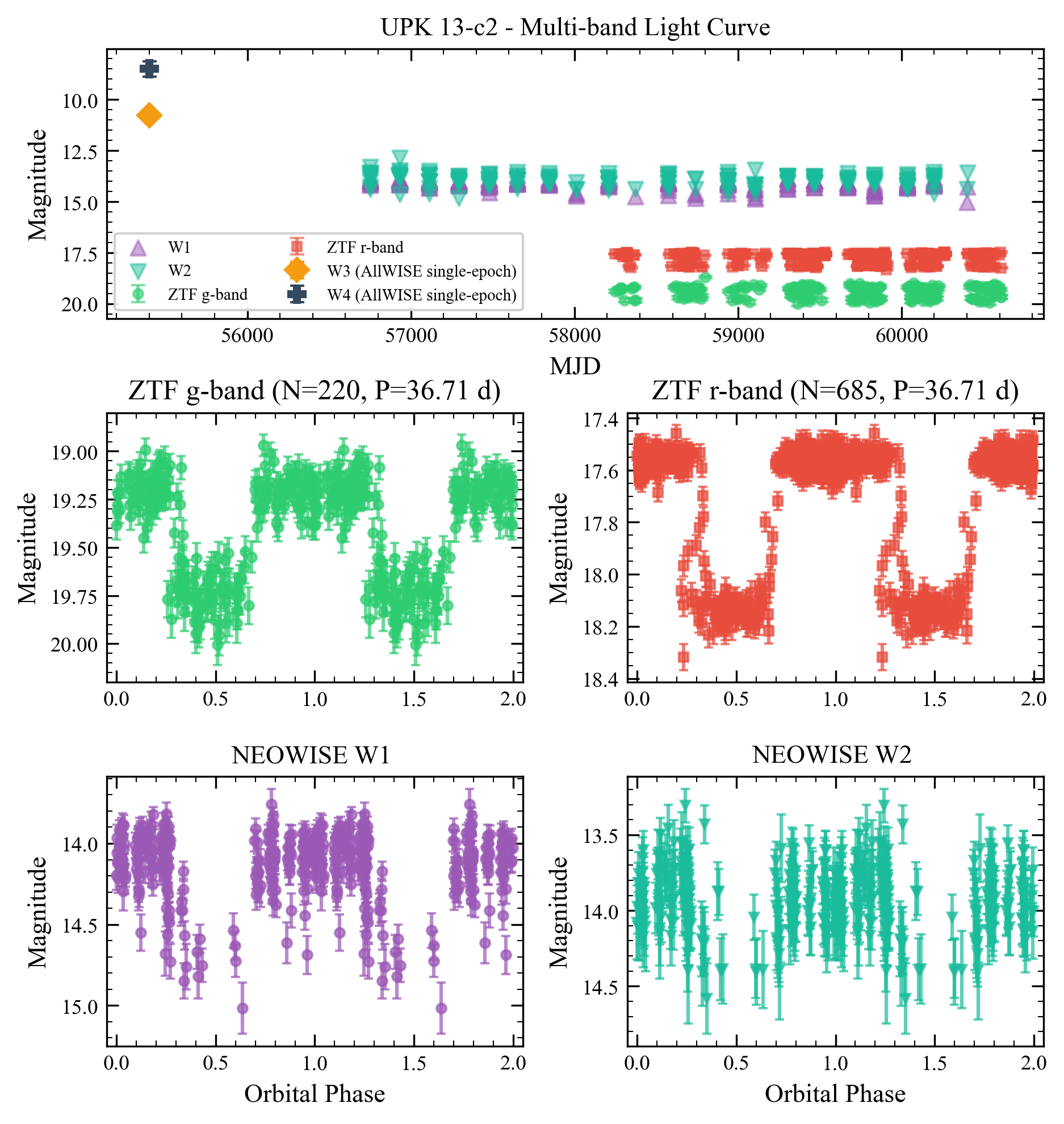}
\caption{Multi-band light curves of UPK~13-c2 phased at $P=36.71$~days.
Top: time series in ZTF $g$, $r$ and NEOWISE $W1$/$W2$;
AllWISE $W3$/$W4$ shown as single error bars at the 2010 epoch.
Middle: phase-folded ZTF $g$ and $r$.
Bottom: phase-folded NEOWISE $W1$ and $W2$.
The AllWISE $W3$/$W4$ points are single-epoch co-added measurements
included to illustrate the mid-infrared excess; they do not constrain
the eclipse profile.}
\label{fig:lightcurve}
\end{figure*}

We retrieved ZTF DR photometry in $g$ ($N=220$) and $r$ ($N=685$)
spanning MJD~$\approx$~58246--61500 (2018--2024). Period analysis via
both BLS \citep{Kovacs2002} and Lomb--Scargle
\citep{Lomb1976, Scargle1982} converges on $P=36.71\pm0.01$~days
($T_0=\mathrm{HJD}~2458246.9421$). Throughout, we adopt the
photometric ephemeris $T(\phi)=T_0+P\,E$ to phase all single-epoch
data. Figure~\ref{fig:lightcurve} shows: (i) a flat out-of-eclipse
baseline ($g\approx 19.2$, $r\approx 17.5$~mag); (ii) sharp
ingress/egress, each $\sim$2--3~days; (iii) a flat-bottomed dimming
of $\sim$10~days with $\Delta g\approx 0.55$ and $\Delta r\approx
0.57$~mag (trapezoidal-fit values).

\subsection{WISE/NEOWISE Mid-Infrared Photometry} \label{subsec:wise}

NEOWISE-R provides $W1$ and $W2$ time-domain photometry over
$\sim$10~years (2014--2024, $\sim$200 single-exposure detections per
band, $\sim$11 epochs of $\sim$1-day visibility windows). The
phase-folded $W1/W2$ light curves show clear eclipses at the same
36.71-day period (Figure~\ref{fig:lightcurve}), demonstrating that
the occulting structure is optically thick at $3$--$5~\mu$m. AllWISE
$W3$ and $W4$ photometry is available only as the
single-epoch co-added cryo-mission measurement (observed in 2010);
no NEOWISE-R or unWISE multi-epoch coverage exists at $W3/W4$. The
quoted catalog uncertainties $\sigma(W3)=0.12$~mag and
$\sigma(W4)=0.38$~mag therefore set the only available upper limit
on phase-coherent variability ($\Delta m\lesssim 0.03$~mag at the
single-coadd level after accounting for the underlying
$\sim$70-frame coadd statistics; \citealt{Cutri2013}). The large
$K_s-W3\approx 3.4$ and $K_s-W4\approx 5.6$~mag excesses confirm a
massive, spatially extended circumstellar dust component.

\subsection{2MASS, Pan-STARRS1, and SPHEREx Photometry}
\label{subsec:nir}

The 2MASS magnitudes ($J=15.115\pm 0.049$, $H=14.332\pm 0.026$,
$K_s=14.133\pm 0.070$~mag) and PS1 photometry ($i=16.951$,
$z=16.587$, $y=16.471$~mag) are single-epoch measurements that fall
in the out-of-eclipse phase window. SPHEREx low-resolution
spectrophotometry over $0.75$--$5~\mu$m (96 channels) independently
validates a late-K/early-M binary composite continuum and reveals the smooth
stellar-to-disk flux transition near $3$--$5~\mu$m; the SPHEREx
data are also dominated by high-state phase coverage and are
therefore used here only as a high-state spectral constraint.

Line-of-sight extinction is accounted for in all SED fits below
using the \citet{Fitzpatrick1999} extinction law with $R_V=3.1$
(extended into the mid-infrared via the empirical
$A(\lambda)/A_V=(\lambda/0.55~\mu{\rm m})^{-1.84}$ relation of
\citet{MartinWhittet1990}), with $A_V=0.85$~mag fixed to the
\citet{Green2019} Bayestar2019 3D dust map value at the source
distance (see Table~\ref{tab:params} and Section~\ref{subsec:sed}).

\begin{deluxetable*}{lcccccc}
\tablecaption{System Parameters of UPK~13-c2 \label{tab:params}}
\tablewidth{0pt}
\tablehead{
\multicolumn{7}{c}{\textit{Astrometric and Classification}} \\
\colhead{Parameter} & \multicolumn{5}{c}{Value} & \colhead{Source}
}
\startdata
R.A. (J2000)              & \multicolumn{5}{c}{$281\fdg0806$}              & Gaia DR3 \\
Decl. (J2000)             & \multicolumn{5}{c}{$-17\fdg8927$}              & Gaia DR3 \\
$\pi$ (mas)               & \multicolumn{5}{c}{$1.013\pm0.132$ (catalog); $1.051\pm0.133$ (adopted)}            & Gaia DR3 ; this work \\
$\mu_\alpha\cos\delta$ (mas\,yr$^{-1}$) & \multicolumn{5}{c}{$-0.62\pm0.14$} & Gaia DR3 \\
$\mu_\delta$ (mas\,yr$^{-1}$) & \multicolumn{5}{c}{$-4.26\pm0.12$}         & Gaia DR3 \\
$G/G_{\rm BP}/G_{\rm RP}$ (mag) & \multicolumn{5}{c}{$17.623/18.810/16.652$} & Gaia DR3 \\
RUWE                      & \multicolumn{5}{c}{$0.977$}                    & Gaia DR3 \\
$P_{\rm WD+MS}$ / Tier    & \multicolumn{5}{c}{$0.9944$ / 2}               & \citet{Grondin2024} \\
$\chi^2_{\rm spat}/\chi^2_{\rm kin}$ & \multicolumn{5}{c}{$42.77/10.57$}   & \citet{Grondin2024} \\
Period (days)             & \multicolumn{5}{c}{$36.71\pm0.01$}             & This work \\
Eclipse duration (days)   & \multicolumn{5}{c}{$\sim 10$}                  & This work \\
Cluster age / distance$^a$ (if member) & \multicolumn{5}{c}{$\sim 316$~Myr / $\sim 951$~pc (adopted)} & \citet{CantatGaudin2020} \\
Adopted $A_V^c$            & \multicolumn{5}{c}{$0.85\pm 0.10$~mag}         & \citet{Green2019} \\
\hline
\multicolumn{7}{c}{\textit{Multi-band Eclipse Photometry}} \\
\hline
Band & $\lambda_{\rm eff}$ ($\mu$m) & $m_{\rm high}$ (mag) & $m_{\rm low}$ (mag) & $\Delta m$ (mag) & $\Delta F/F_{\rm high}$ & Source \\
\hline
ZTF $g$           & $0.472$ & $19.19$ & $19.74$ & $0.55$ & $0.40$ & ZTF \\
ZTF $r$           & $0.634$ & $17.56$ & $18.13$ & $0.57$ & $0.41$ & ZTF \\
PS1 $i$           & $0.752$ & $16.95$ & \nodata & \nodata & \nodata & Pan-STARRS1 \\
PS1 $z$           & $0.866$ & $16.59$ & \nodata & \nodata & \nodata & Pan-STARRS1 \\
PS1 $y$           & $0.962$ & $16.47$ & \nodata & \nodata & \nodata & Pan-STARRS1 \\
2MASS $J$         & $1.235$ & $15.12$ & \nodata & \nodata & \nodata & 2MASS \\
2MASS $H$         & $1.662$ & $14.33$ & \nodata & \nodata & \nodata & 2MASS \\
2MASS $K_s$       & $2.159$ & $14.13$ & \nodata & \nodata & \nodata & 2MASS \\
NEOWISE $W1$      & $3.35$  & $14.06$ & $14.68$ & $0.62$ & $0.44$ & NEOWISE \\
NEOWISE $W2$      & $4.60$  & $13.93$ & $14.28$ & $0.35$ & $0.28$ & NEOWISE \\
AllWISE $W3$$^b$  & $11.56$ & $10.78$ & \nodata & $<0.03^d$ & $<0.03^d$ & AllWISE \\
AllWISE $W4$$^b$  & $22.09$ & $8.50$  & \nodata & $<0.03^d$ & $<0.03^d$ & AllWISE \\
\enddata
\tablecomments{Eclipse magnitudes from trapezoidal fits to phase-resolved
ZTF and NEOWISE light curves. $\Delta F/F_{\rm high}=1-10^{-\Delta m/2.5}$.
$^a$Cluster parameters from \citet{CantatGaudin2020}; UPK~13-c2 membership is
not yet spectroscopically confirmed. $^b$AllWISE provides only co-added
cryo-mission photometry (single epoch, observed 2010); the listed
magnitude is therefore not phase-resolved.
$^c$Extinction is fixed to the Bayestar2019 three-dimensional dust map
value \citep{Green2019} at the source distance; the assigned
$\pm 0.10$~mag brackets the dust-map quoted uncertainty at this
sightline (see Section~\ref{subsec:sed}).
$^d$Observed upper limit set by the AllWISE single-coadd precision.}
\end{deluxetable*}

%% ======================================================================
\section{Analysis} \label{sec:analysis}

\subsection{Light Curve Analysis} \label{subsec:lc}

The square-wave morphology---flat baseline, sharp transitions, flat
floor---is the geometric signature of an opaque screen sweeping
across one component of a binary
\citep{Winn2004, Winn2006, Zhu2022, Hu2024, Hu2026, Poon2021}. The
fractional flux decrements are approximately achromatic across the
optical-to-$W1$ window ($\Delta F/F_{\rm high}\approx 0.40$--$0.44$,
mean $\approx 0.42$), with $W2$ reduced to $\sim$0.28 because of
disk thermal dilution at $4.6~\mu$m. The flat floor requires the \emph{complete} occultation of one stellar component. We quantify this geometric requirement: the
observed ingress duration of $\sim$2.5~days, combined with the
projected stellar radius, requires the disk-edge gradient to span
$\lesssim 3 R_\star\sim 1.8\,R_\odot$ at the chord crossing, i.e., a
geometrically sharp truncation.

\subsection{SED Decomposition: WD$+$MS vs.\ MS$+$MS} \label{subsec:sed}

We construct the broadband SED using the Gaia, PS1, 2MASS, NEOWISE,
AllWISE, and SPHEREx photometry (12 broadband points $+$ 7 binned
SPHEREx channels $=$ 19 spectral constraints).
All photometry is dereddened with the Bayestar2019 three-dimensional
dust map value at the source distance, $A_V=0.85$~mag
\citep{Green2019}, fixed throughout the analysis.
The SED extends from 0.4 to 22~$\mu$m, with the 5--22~$\mu$m regime
dominated by a prominent mid-infrared excess at $W3/W4$ requiring
a circumstellar dust component.
To represent the underlying stellar photospheres, we utilize the
BaSeL (v2.2) empirical stellar spectra library
\citep{Lejeune1997,Lejeune1998} via the \texttt{pystellibs} package,
generating continuous spectra that match our photometrically derived
parameters. We caution that the specific
subtype labels (M1V/K9V) returned by the broadband template
matching should not be overinterpreted: they are template-based
descriptors of the photometric color and slope, and the physically
robust conclusion is that the primary and the occulted component sit
near the late-K/early-M boundary, rather than a precise spectral
classification.

The flat-bottomed eclipse guarantees that the low-state flux contains
zero contribution from the eclipsed component:
$F_{\rm high}=F_{\rm low}+F_{\rm diff}$, where
$F_{\rm diff}\equiv F_{\rm high}-F_{\rm low}$ is the SED of the
occulted body alone. This additive decomposition is exact for
complete occultation. We exploit it identically under both
hypotheses---fitting $F_{\rm diff}$, $F_{\rm low}$, and their sum
$F_{\rm high}$ in three steps---so that any residual pattern is
driven by the physics of the templates, not by a methodological
asymmetry. Throughout this paper we enforce a 20\% systematic floor on
$\sigma_i$ to account for template-to-photometry
systematics and avoid bright-band data dominating the statistic.
2MASS bands receive $4\times$ weight to anchor the near-infrared
photospheric peak.

\paragraph{Hypothesis A: WD$+$MS$+$disk.}
\emph{Step~1 (difference spectrum):} Fitting $F_{\rm diff}$ with
Koester DA model atmospheres \citep{Koester2010} over the grid
$T_{\rm WD}\in[5000,80000]$~K, $\log g\in[6.5,9.5]$
yields a best match at the lower temperature boundary ($T_{\rm WD}=5000$~K, $\log g=9.0$), but with
$\chi^2_{\rm diff}\approx 45$ (4~dof)---a poor fit that strongly
disfavors the WD on the difference spectrum alone
(Section~\ref{subsec:reject}). We restrict the temperature grid to $\ge 5000$~K because a cooler white dwarf, given its intrinsically tiny radius, would have a bolometric luminosity far too low to account for the large $\sim$40\% observed flux decrement. Furthermore, the Rayleigh--Jeans spectral slope of the WD cannot simultaneously match the optical eclipse depth and the large $W1/W2$ decrement.

\begin{figure*}[t!]
\centering
\includegraphics[width=0.95\textwidth]{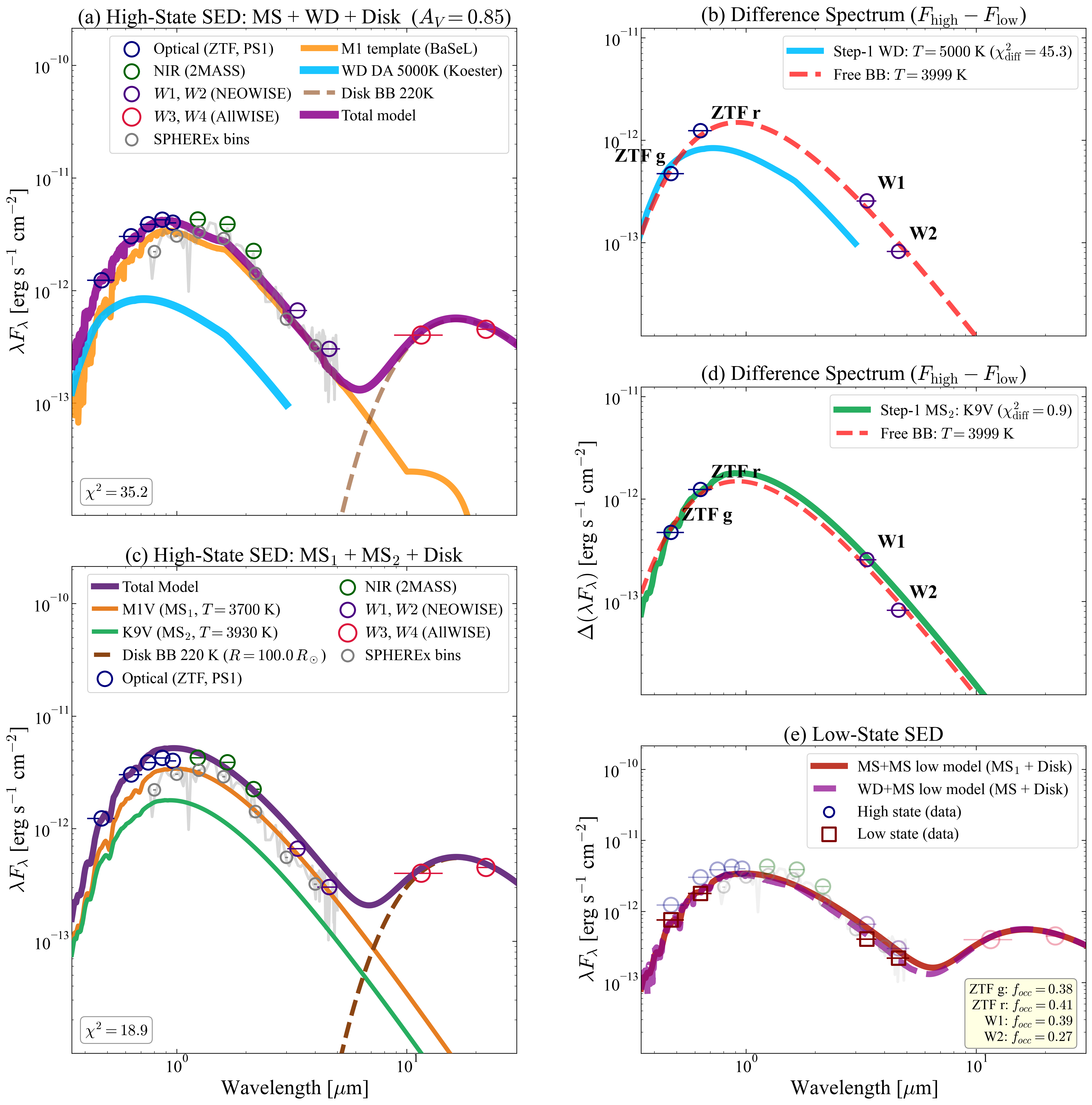}
\caption{SED analysis comparing WD$+$MS (top) and MS$+$MS (bottom)
hypotheses (Section~\ref{subsec:sed}; $A_V=0.85$~mag; $d=951$~pc).
(a)~WD$+$MS high-state SED ($\chi^2=35.25$).
(b)~Difference spectrum: WD strongly disfavored
($\chi^2_{\rm diff}\approx 45$); blackbody reference at
$T\approx 4000$~K (red dashed).
(c)~MS$+$MS high-state SED ($\chi^2=18.88$).
(d)~Difference spectrum: K9V template fits with
$\chi^2_{\rm diff}\approx 0.9$.
(e)~Low-state SED; the two models are nearly
indistinguishable, confirming the discriminating power resides
in the difference spectrum.
The extent of each photometric point along the wavelength axis
indicates the corresponding filter bandwidth. Out-of-eclipse and
in-eclipse observations are denoted by open circles and open squares,
respectively.}
\label{fig:sed}
\end{figure*}

\emph{Step~2 (low state):} Fitting $F_{\rm low}$ with an
early-M stellar template plus a disk blackbody yields an M1V primary with
$T_{\rm disk}\approx 220$~K.
\emph{Step~3 (high-state consistency):} The synthetic
$F_{\rm synth}=F_{\rm low,model}+F_{\rm diff,model}$ yields
$\chi^2=35.25$ on the high-state data (19 bands), because
the WD template absorbs the dereddened blue flux
(Figure~\ref{fig:sed}a,b). However, this global fit
is artificially driven by the broad-band integration, masking the fact
that the WD template cannot naturally reproduce the observed
difference spectrum.

\paragraph{Hypothesis B: MS$+$MS$+$disk.}
\emph{Step~1 (difference spectrum):} Fitting $F_{\rm diff}$ with a
single reddened late-K stellar template yields a K9V at
$T_{\rm eff}\approx 3930$~K (color temperature
$T_{\rm occ}\approx 4000$~K), with
$\chi^2_{\rm diff}\approx 0.9$ (4~dof)---an excellent fit.
\emph{Step~2 (low state):} Fitting $F_{\rm low}$ with a reddened M1V
template ($T_{\rm eff}\approx 3700$~K) plus a disk blackbody
($T_{\rm disk}\approx 220$~K) reproduces the NIR photospheric peak
and the $W3/W4$ excess.
\emph{Step~3 (high-state consistency):} The synthesized high-state
SED yields $\chi^2=18.88$ (19 bands), significantly lower than the
WD$+$MS model ($\chi^2=35.25$), reflecting the MS$+$MS template's
superior match to the broadband photospheric shape from 0.4 to
22~$\mu$m (Figure~\ref{fig:sed}c).
The residual optical offset ($\sim$20\% at $g$, $r$)
is within the systematic uncertainty of the template-matching
procedure: the decoupled three-step method fits $F_{\rm diff}$ and
$F_{\rm low}$ independently, and their sum is not re-optimized on
$F_{\rm high}$, so template-to-photometry mismatches at wavelengths
not constrained by the low-state or difference-spectrum bands
propagate directly into the forward prediction. This residual is
also consistent with the residual systematics of empirical template
matching at optical wavelengths for magnetically active, rapidly
rotating late-type dwarfs \citep{West2008, Kiman2021}, and does not
affect the physical conclusions drawn from the difference spectrum
(Figure~\ref{fig:sed}d).

\paragraph{Summary of hypothesis comparison.}
The identical three-step procedure yields a pronounced contrast on
two levels. First, the difference spectrum: the K9V template fits
with $\chi^2_{\rm diff}=0.9$, while the best WD fit gives
$\chi^2_{\rm diff}=45$. Since the difference spectrum is an
\emph{extinction-independent} observable (the same foreground column
cancels in the subtraction), this discrimination does not depend on
the adopted $A_V$. Second, the high-state SED: the MS$+$MS model
achieves $\chi^2=18.88$ vs.\ $\chi^2=35.25$ for the WD$+$MS model
(Figure~\ref{fig:sed}a,c).
This confirms that the MS$+$MS hypothesis is
the better global fit across all 19 bands. We therefore adopt the
MS$+$MS$+$disk decomposition as the working model.

\paragraph{Binary parameter estimates.}
The mass ratio is constrained by an extinction-independent observable:
the mean depth $\Delta F/F\approx 0.42$ fixes $L_2/L_{\rm tot}\approx 0.42$,
and the empirical mass--luminosity relation of \citet{Mann2019}
gives $q\equiv M_2/M_1\approx 1.04$.
The individual stellar masses are inferred by integrating the
best-fit SED templates, scaled to absolute flux at $d\approx 951$~pc, to obtain the absolute
bolometric luminosities ($L_1\approx 0.087\,L_\odot$,
$L_2\approx 0.103\,L_\odot$, $L_{\rm tot}\approx 0.19\,L_\odot$),
which are then converted to masses via empirical mass--luminosity
relations \citep{Mann2019}.
The best-fit BaSeL template labels are M1V for the low-state primary
($T_1\approx 3700$~K) and K9V for the occulted component
($T_2\approx 3930$~K). Because broadband template labels near the
late-K/early-M boundary are sensitive to extinction and template
systematics, we treat these as color descriptors rather than precise
spectral classifications. Scaling the SED luminosities and applying
empirical mass--luminosity relations yields photometrically estimated
$M_1\approx 0.68\,M_\odot$, $M_2\approx 0.70\,M_\odot$, and
$M_{\rm tot}\approx 1.4\,M_\odot$.
These derived fundamental parameters carry ${\sim}20\%$ systematic
uncertainties from broadband template matching; precise values await
high-resolution spectroscopic or dynamical mass constraints.
The dereddened Gaia color
$(G_{\rm BP}-G_{\rm RP})_0\approx 1.3$~mag of the integrated
photometry is bluer than the single late-type dwarf locus,
consistent with the composite spectrum of the binary.

\subsection{Disk Geometry} \label{subsec:geometry}

\begin{figure*}[t!]
\centering
\includegraphics[width=0.95\textwidth]{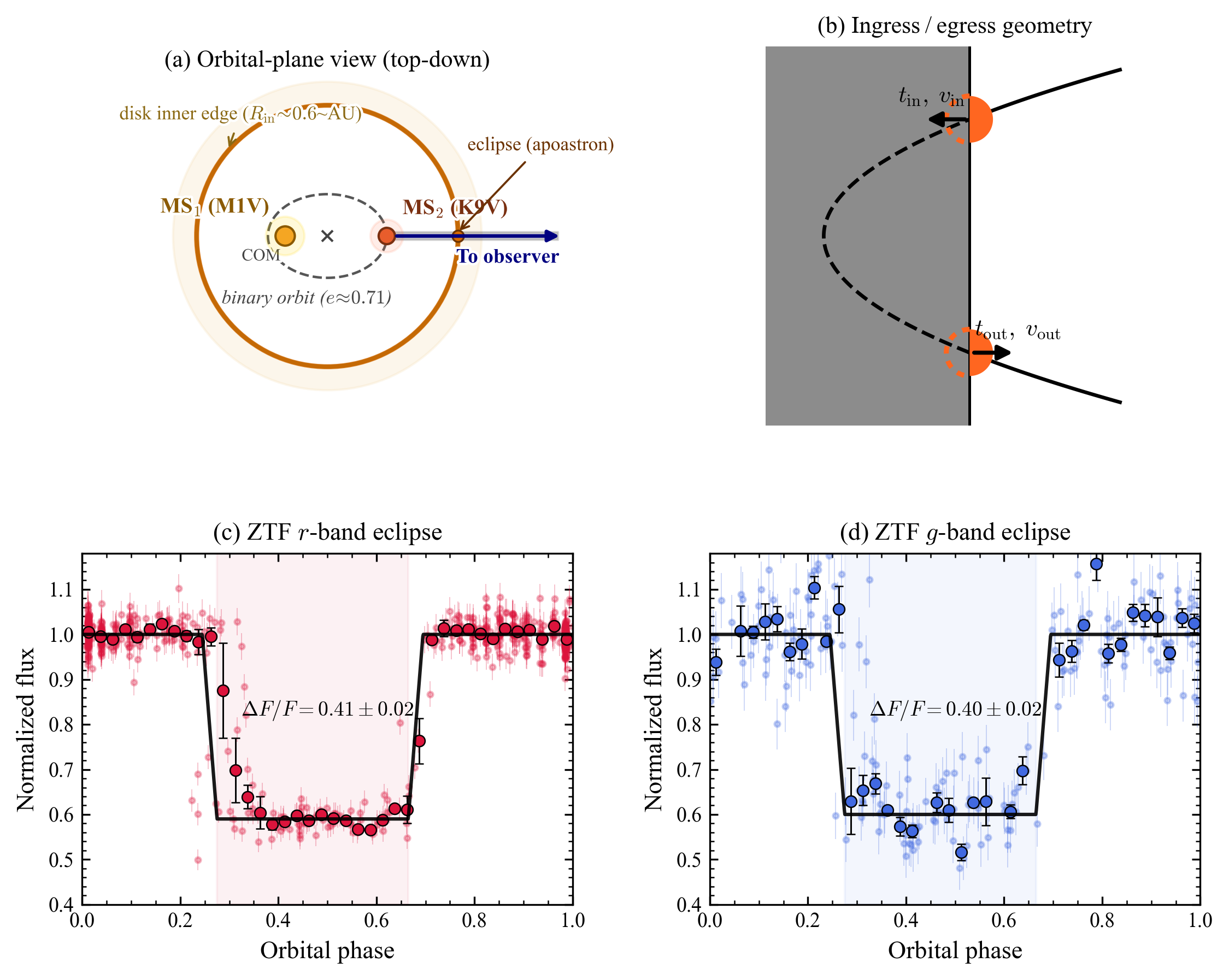}
\caption{Disk-occultation geometry of UPK~13-c2.
(a)~Top-down orbital-plane view showing the eccentric binary orbit
($e\!\approx\!0.71$) with the tidally-truncated disk inner edge
($R_{\rm in}\!\sim\!0.5$--$0.7$~AU).
(b)~Sky-plane ingress/egress geometry.
(c)~Phase-folded ZTF $r$-band light curve with trapezoidal model
($\Delta F_r/F=0.41$).
(d)~Phase-folded ZTF $g$-band light curve ($\Delta F_g/F=0.40$);
near-identical $g$/$r$ depths confirm achromatic occultation.}
\label{fig:diskmodel}
\end{figure*}

For $M_{\rm tot}\approx 1.4\,M_\odot$ and $P=36.71$~days, Kepler's
third law gives a semi-major axis
\begin{equation}
a = \left(\frac{GM_{\rm tot}P^2}{4\pi^2}\right)^{1/3} \approx 0.24~{\rm AU}
   \approx 52\,R_\odot\,,
\end{equation}
with mean orbital velocity $v_{\rm orb}\approx 71$~km\,s$^{-1}$.
Tidal truncation \citep{Artymowicz1994} sets the circumbinary inner
edge at $R_{\rm in}\sim 2$--$3\,a\approx 0.48$--$0.72$~AU, where the
equilibrium dust temperature for $L_{\rm tot}\approx 0.19\,L_\odot$
is $T_{\rm in}\sim 220$--$270$~K---in excellent agreement with the
$\sim$220~K independently inferred from the $W3/W4$ excess (which
corresponds to a radiative equilibrium radius $R_{\rm eq}\approx 0.72$~AU,
fully consistent with the tidal truncation range). At this
radius a disk scale height $H/R\sim 0.1$--$0.15$ gives
$H\sim 15$--$22\,R_\odot$, easily sufficient to occult a
$\sim$0.70~$R_\odot$ late-K dwarf at the binary semi-major axis. The
geometric configuration sketched in Figures~\ref{fig:diskmodel}a,b
naturally produces the flat-bottomed eclipses seen in both ZTF
$r$ (Figure~\ref{fig:diskmodel}c) and ZTF $g$
(Figure~\ref{fig:diskmodel}d).

The ingress/egress duration ($\sim$2.5~days) implies a perpendicular
crossing velocity
$v_\perp = 2R_2/t_{\rm ingress}\approx 4.5$~km\,s$^{-1}$, a factor
of $\sim$16 slower than $v_{\rm orb}$. This discrepancy resolves
naturally as the combination of two effects: (i) a grazing
disk-edge geometry, $v_\perp = v_{\rm orb}\sin\alpha$, which permits
small grazing angles $\alpha$, and (ii) an eccentric orbit with
eclipse near apoastron, where
$v_{\rm apo}=v_{\rm orb}\sqrt{(1-e)/(1+e)}$ is reduced by a factor
$\sim$2--3 for $e\sim 0.5$--$0.7$. Both effects are simultaneously
constrained by Kepler's second law applied to the observed
fractional eclipse duration $\tau/P\approx 0.27$:
\begin{equation}
\frac{\tau}{P} = \frac{\alpha}{\pi}\,
                 \frac{(1-e^2)^{3/2}}{(1+e\cos\nu_{\rm mid})^{2}}\,,
\label{eq:kepler2}
\end{equation}
where $\nu_{\rm mid}$ is the true anomaly at mid-eclipse and $2\alpha$
is the disk azimuthal arc subtended along the line of sight; $\tau$
denotes the total eclipse duration (i.e., the full time between first
and last contact).

To explore the eclipse geometry quantitatively, we fit a
3-parameter ($e$, $\omega$, $\alpha$) forward eclipse
model using MCMC (\textsc{emcee}, 64 walkers, $2\,000$ burn $+$
$4\,000$ production steps). The fractional depth
$\Delta F/F\approx 0.41$ is not a fit parameter: it is set by the
secondary-to-total luminosity ratio
$\varepsilon\equiv L_2/L_{\rm tot}$ under complete occultation and
is fixed directly by the observed eclipse amplitude (see
Section~\ref{subsec:sed}), so including it in the likelihood would
amount to a tautological recovery of its input value. The
morphological MCMC therefore constrains only the three
shape parameters $(e,\omega,\alpha)$, which jointly determine the
eclipse duration and ingress fraction; $\varepsilon$ enters
separately as a luminosity-ratio constraint on the binary
components.

Our eclipse model assumes a sharply truncated (knife-edge) disk boundary, where the opacity transitions from zero to unity over a radial scale $\ll R_\star$. Under this assumption, the ingress duration $t_{\rm ingress}$ is governed entirely by the geometric chord-crossing time ($\propto R_\star/v_\perp$). However, if the disk edge features a finite optical-depth gradient---as demonstrated for KH~15D \citep{Winn2004, Winn2006}---the observed ingress will partly reflect the spatial scale of this opacity profile. Consequently, the derived values of $e$ and $\alpha$ are partially degenerate with the unconstrained edge structure. The parameters reported below should therefore be treated as morphological fits under a sharp-edge prior rather than definitive kinematic constraints. A robust determination of the orbital eccentricity awaits multi-epoch radial velocity (RV) follow-up of the primary, analogous to the spectroscopic confirmation of Bernhard-2 \citep[$e=0.69$;][]{Hu2024}.

We further note a mild internal tension that highlights this
degeneracy. Substituting the posterior medians $e=0.71$ and
$\alpha=13\fdg3$ into the apoastron projection
$v=v_{\rm orb}\sqrt{(1-e)/(1+e)}\,\sin\alpha$ yields
$v\approx 6.7$~km\,s$^{-1}$, which for $R_2\approx 0.70\,R_\odot$
predicts an ingress of $\sim$1.7~days, slightly shorter than the
observed $\sim$2.5~days. This $\sim$30\% mismatch is the expected
signature of a finite optical-depth gradient at the disk edge: when
the sharp-edge prior absorbs all of the ingress timescale into
geometric chord crossing, the fit underestimates the kinematic
ingress duration. A more realistic edge model with non-zero opacity
scale would relax this tension by re-distributing part of the ingress
time onto the opacity profile, allowing slightly larger $\alpha$ or
smaller $e$. We therefore reiterate that the $(e,\alpha)$ values are
shape descriptors, and quantitative reconciliation requires both RV
constraints and an opacity-gradient model along the lines of
\citet{Winn2004, Winn2006}.

\begin{figure}[t!]
\centering
\includegraphics[width=\columnwidth]{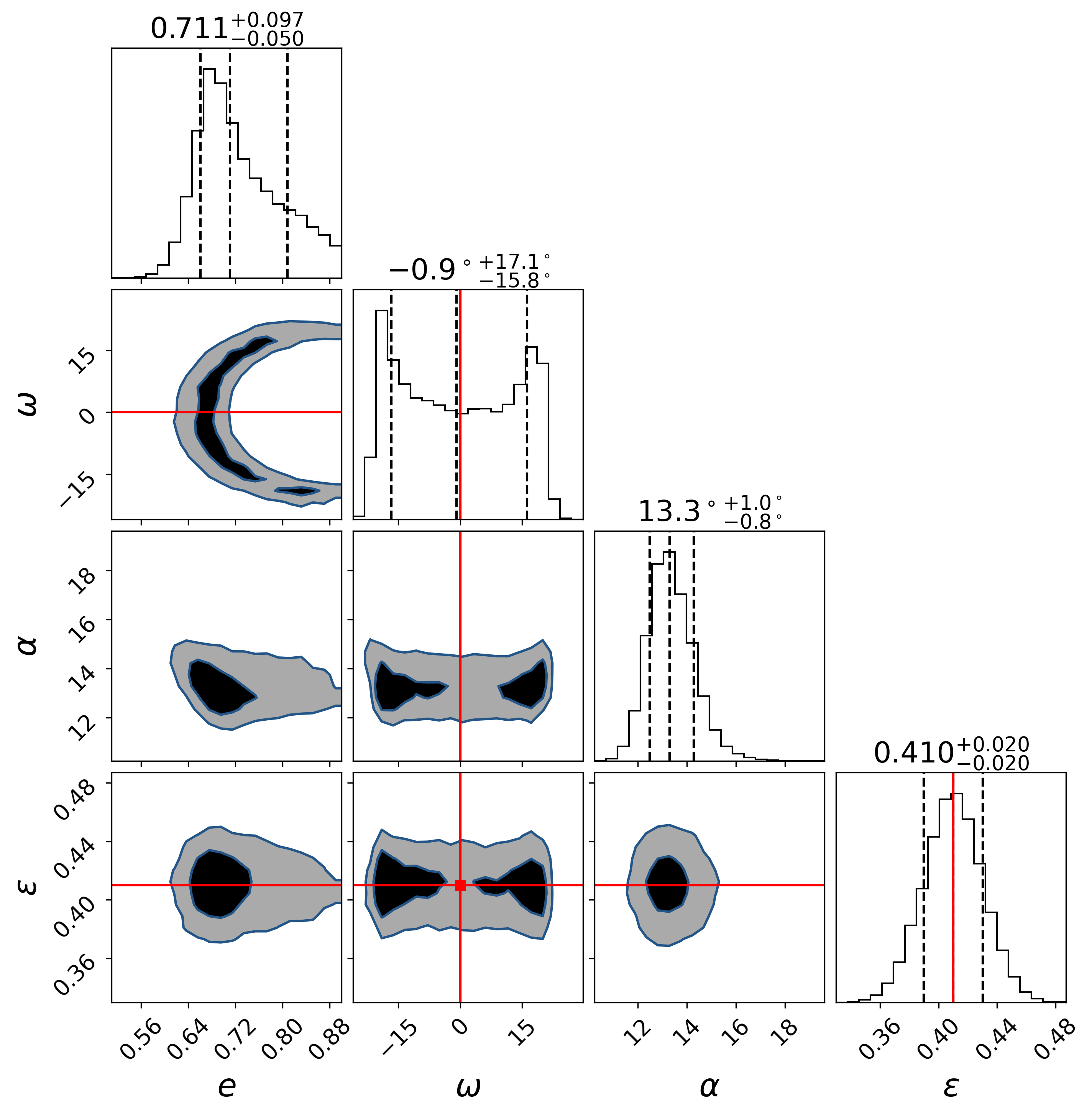}
\caption{Posterior distributions for the three-parameter
$(e,\omega,\alpha)$ forward eclipse model
(Section~\ref{subsec:geometry}); the luminosity-ratio parameter
$\varepsilon=\Delta F/F=0.41\pm 0.02$ is fixed externally by the
observed eclipse depth and shown for reference only. Contours show
1$\sigma$ and 2$\sigma$ levels; titles report median and
16/84-percentile bounds. Red crosshairs mark the reference values
$\omega=0$ and $\varepsilon=0.41$. The inferred $e$ is a
morphology-based descriptor under the sharp-edge prior and is not
a direct kinematic eccentricity measurement.}
\label{fig:mcmc}
\end{figure}

With this limitation in mind, the MCMC uses Gaussian likelihoods on
the two morphological observables
($\tau/P=0.272\pm 0.014$, ingress$/P=0.068\pm 0.008$),
a soft Gaussian prior centered on
$\omega=0$ with $\sigma_\omega=\pi/4$ enforcing the symmetric
apoastron-eclipse geometry, and uniform priors on $e$ and $\alpha$.
The posterior is shown in Figure~\ref{fig:mcmc}; the
best-fit morphological parameters are
\begin{align}
e &= 0.71^{+0.10}_{-0.05}, \\
\omega &= -0\fdg9^{+17^\circ}_{-16^\circ} \quad \text{(consistent with apoastron)}, \\
\alpha &= 13\fdg3^{+1.0}_{-0.8}.
\end{align}
The luminosity-ratio parameter is set externally by the observed
depth, $\varepsilon=\Delta F/F=0.41\pm 0.02$.
The $(e,\alpha)$ pair is anti-correlated by
Eq.~\ref{eq:kepler2}: a more eccentric orbit slows the apoastron
crossing, so a smaller arc $\alpha$ matches the same $\tau/P$. The
inferred $\alpha\approx 13^\circ$ corresponds to an arc length of
$\sim$$0.2$~AU at $R_{\rm in}\sim 0.5$--$0.7$~AU, consistent with a
localized misaligned ring rather than a near-complete azimuthal
opacity blanket. The depth-implied chord through the disk at
projected radius $R_{\rm in}\approx 0.5$--$0.7$~AU together with
$\alpha\approx 13^\circ$ gives a moderate-to-high viewing inclination
$i_{\rm disk}\sim 60^\circ$--$80^\circ$; the absolute value of
$i_{\rm disk}$ remains weakly constrained without RV measurements.
These values are similar in magnitude to the RV-confirmed eccentricities of KH~15D ($e=0.68$) and Bernhard-2 ($e=0.69$); however, the UPK~13-c2 estimate represents a photometric shape descriptor rather than a direct dynamical constraint.

%% ======================================================================
\section{Discussion} \label{sec:discussion}

\subsection{Independent Diagnostics Disfavoring the WD$+$MS Hypothesis}
\label{subsec:reject}

As shown in Section~\ref{subsec:sed}, the identical three-step
decoupled-fit procedure yields a pronounced contrast at the level of
$F_{\rm diff}$ ($\chi^2_{\rm diff}\approx 45$ for the WD vs.\ $0.9$
for K9V), while the global high-state SED alone is less
discriminating between the two hypotheses
($\chi^2=35.25$ vs.\ $18.88$). We therefore
base the argument on two diagnostics that are independent of the
global SED partition: the eclipse morphology and the spectral shape
of the differenced flux. The strongest evidence comes from the
eclipse morphology and the difference spectrum; the global
high-state SED fit is a secondary, less discriminating cross-check.

\paragraph{(1) Flat-bottomed eclipse geometry strongly disfavors a WD as the
occulted component.}
The flat-bottomed, square-wave morphology
(Figure~\ref{fig:diskmodel}c,d) requires \emph{complete} occultation
of a single component during mid-eclipse, as established in KH~15D
\citep{Winn2004, Winn2006, Poon2021} and Bernhard-2
\citep{Hu2024, Hu2026}. Partial occultation of a single star on a
curved orbit produces a continuously varying covering fraction and
hence a U-shaped profile; a flat floor lasting $\sim$10~days
demands the occulted body remain entirely hidden across that interval.
Quantitatively, the observed ingress of $\sim$2.5~days requires the
occulter edge gradient to span $\lesssim 3R_\star$, i.e.\ to be
geometrically sharp on a scale comparable to the occulted stellar
diameter. The radius of an $\sim$0.6~$M_\odot$ DA white dwarf is
$R_{\rm WD}\!\sim\!0.013\,R_\odot\!\sim\!9{,}000$~km---some
$54\times$ smaller than the secondary late-K dwarf
($R_2\!\approx\!0.70\,R_\odot$). For the WD to be the body that
the disk hides, the disk-edge crossing of the WD would have to last
\emph{at most} $t_{\rm cross}^{\rm WD}=t_{\rm ingress}\times
(R_{\rm WD}/R_\star)\!\sim\!2$~hours, producing a deep but
needle-narrow dip---completely incompatible with the smooth $\sim$2.5~d
ingress and 10~d flat floor that is observed. Conversely, occultation
of the K9V companion by a circumbinary disk reproduces both
timescales naturally because $R_2$ matches the inferred $\sim
3R_\star$ disk-edge gradient by construction. The flat-bottomed
geometry therefore strongly disfavors a scenario in which the WD is
the body being eclipsed.

\paragraph{(2) Strong statistical evidence against the WD hypothesis from the difference spectrum.}
Under the decoupled-fit framework, and under the complete-occultation
assumption, the differenced component
$\Delta F_\lambda=F_\lambda^{\rm high}-F_\lambda^{\rm low}$ contains
no contribution from the primary; it is produced entirely by the
eclipsed component. A K9V template at $T_{\rm eff}\approx
3930$~K (color temperature $\sim$4000~K) fits $F_{\rm diff}$ with
$\chi^2_{\rm diff}\approx 0.9$ (4~dof). A white dwarf, by contrast,
cannot naturally reproduce $F_{\rm diff}$: the WD
spectral energy distribution follows a steep Rayleigh--Jeans tail
($F_\nu\propto\nu^2$) longward of its Wien peak at
$\lambda\lesssim 0.58~\mu$m ($T_{\rm WD}\approx 5000$~K). To match
the observed optical eclipse depths in $g$ and $r$, the WD must
supply $\sim$40\% of the broadband flux at 0.5--0.6~$\mu$m; yet
the same Rayleigh--Jeans slope then plunges by orders of magnitude
toward $W1$ (3.4~$\mu$m) and $W2$ (4.6~$\mu$m), yielding
predicted infrared eclipse depths $\lesssim$1\%---a factor of
$\sim$40 below the observed $\Delta F/F(W1)\approx 0.44$ and
$\sim$28 below $\Delta F/F(W2)\approx 0.28$. A forced WD fit to
$F_{\rm diff}$ yields $\chi^2_{\rm diff}\approx 45$ even for the
best-matching 5000~K model, providing strong evidence against
the WD hypothesis. If the WD luminosity is
artificially scaled up to match the infrared eclipse depths, its
blue-end flux would far exceed the observed high-state
$g/r$ photometry.

\subsection{The MS$+$MS$+$Disk Solution}
\label{subsec:msms}

An inferred late-K/early-M binary (photometrically estimated M1V$+$K9V;
template-based $M_1\approx 0.68\,M_\odot$, $M_2\approx 0.70\,M_\odot$,
$M_{\rm tot}\approx 1.4\,M_\odot$, derived from SED
bolometric luminosities via \citealt{Mann2019}) with a misaligned
circumbinary disk simultaneously explains the flat-bottomed eclipse,
the approximately achromatic $\sim$40\% flux decrement from optical
through $W1$, the $\sim$4000~K difference-spectrum color temperature,
and the $\sim$220~K dust temperature inferred from the $W3/W4$ excess
(matching the equilibrium temperature at the tidally truncated
inner rim). The MS$+$MS model provides the better global fit
($\chi^2=18.88$ vs.\ $35.25$ for WD$+$MS; Section~\ref{subsec:sed}). The same geometry is observed in
KH~15D and Bernhard-2, with UPK~13-c2 providing a late-K/early-M dwarf
analogue at substantially lower luminosity ratio (and hence
substantially shallower eclipse depth). The slightly larger
$W1$ decrement compared to $W2$ ($0.44$ vs.\ $0.28$) is
quantitatively explained by disk thermal dilution at $4.6~\mu$m.

\subsection{Comparison with Known Disk-Eclipsing Systems}
\label{subsec:comparison}

UPK~13-c2 shares the defining properties of KH~15D-class
disk-eclipsing binaries (Table~\ref{tab:comparison}). Its eclipse
depth ($\sim$0.5--0.6~mag) is much shallower than KH~15D
($\sim$3.5~mag) or Bernhard-2 ($\sim$1.5~mag). In KH~15D the deep
eclipse is set by the time-dependent ring-edge geometry that
\emph{fully} occults the more luminous primary (component~B), so
``contrast'' there refers to the visibility of the fainter A--B
luminosity ratio rather than to a single mass ratio.
For UPK~13-c2 the depth is set primarily by the
secondary-to-total luminosity ratio of two cool dwarfs of comparable
mass ($L_2/L_{\rm tot}\approx 0.42$, $q\approx 1.04$), which is a
much smaller contrast than in either KH~15D or Bernhard-2 and
naturally explains the shallower flat floor.

\begin{deluxetable*}{lccc}
\tablecaption{Comparison with Known Disk-Eclipsing Binaries
\label{tab:comparison}}
\tablewidth{0pt}
\tablehead{
\colhead{Property} & \colhead{UPK~13-c2} & \colhead{KH~15D} & \colhead{Bernhard-2}
}
\startdata
Period (days)         & $36.71$            & $48.37$            & $63.36$ \\
Eccentricity (morph.)$^{b}$  & $0.71^{+0.10}_{-0.05}$ (this work)$^{b}$ & $0.68\pm0.03$ & $0.69\pm0.08$ \\
$M_1/M_2$ ($M_\odot$) & ${\sim}0.68/{\sim}0.70$$^c$        & $0.6/0.5$          & $1.1/0.9$ \\
$q=M_2/M_1$           & $\sim 1.04$        & $\sim 0.83$        & $\sim 0.82$ \\
Eclipse depth (mag)   & $0.5$--$0.6$       & $\sim 3.5$         & $\sim 1.5$ \\
Eclipse fraction      & $\sim 0.27$        & $0.40$--$0.50$     & $\sim 0.50$ \\
Eclipse morphology    & Flat-bottomed      & Flat-bottomed      & Flat-bottomed \\
Age (Myr)             & $\sim 316^{a}$ (if confirmed) & $\sim 3$           & $\lesssim 20$ \\
Binary type           & MS$+$MS (proposed) & T~Tau$+$T~Tau      & MS$+$MS (confirmed) \\
Disk type             & Circumbinary       & Circumbinary       & Circumbinary \\
MIR excess            & $W3$, $W4$         & Yes                & Yes \\
Spec.\ confirmation   & Pending            & Yes                & Yes \\
\enddata
\tablecomments{KH~15D parameters from
\citet{Herbst2002, Winn2006, Chiang2004, Poon2021}; Bernhard-2 from
\citet{Zhu2022, Hu2024, Hu2026}. UPK~13-c2's MS$+$MS classification
is proposed in this work and awaits spectroscopic confirmation.
$^a$Age applies if cluster membership is spectroscopically
confirmed.
$^b$Morphological descriptor from the 3-parameter
$(e,\omega,\alpha)$ forward eclipse-model MCMC under
the sharp-edge prior (Section~\ref{subsec:geometry};
Figure~\ref{fig:mcmc}); not a direct kinematic measurement.
The UPK~13-c2 value is a light-curve-shape descriptor and is not
directly comparable to the RV-confirmed eccentricities of KH~15D
and Bernhard-2; RV confirmation is required before any
quantitative comparison.
$^c$Masses estimated from SED bolometric luminosities via
\citet{PecautMamajek2013} and \citet{Mann2019};
uncertainty ${\sim}20\%$. Dynamical masses await RV follow-up.}
\end{deluxetable*}

\subsection{Disk Longevity to $\sim$300~Myr}
\label{subsec:longevity}

If cluster membership is confirmed, the $\sim$316~Myr age would make
UPK~13-c2 possibly the longest-lived occulting disk system yet
detected around a main-sequence binary, exceeding Peter~Pan disk
lifetimes \citep{Silverberg2020} by nearly an order of magnitude.
This system raises the possibility that such obscuring circumbinary
material can survive to substantially older ages than in currently
confirmed examples. We note that the present photometry provides no direct constraints on the physical properties or evolutionary origin of the disk. Current data cannot distinguish between a long-lived primordial reservoir preserved by binary tidal truncation \citep{Artymowicz1994, Lai2023} and a massive, optically thick second-generation debris disk. Consequently, we leave both as open possibilities, pending future spectroscopic and interferometric observations.

\subsection{Cluster Membership of UPK~13-c2}
\label{subsec:dynorigin}

\begin{figure*}[t!]
\centering
\includegraphics[width=0.95\textwidth]{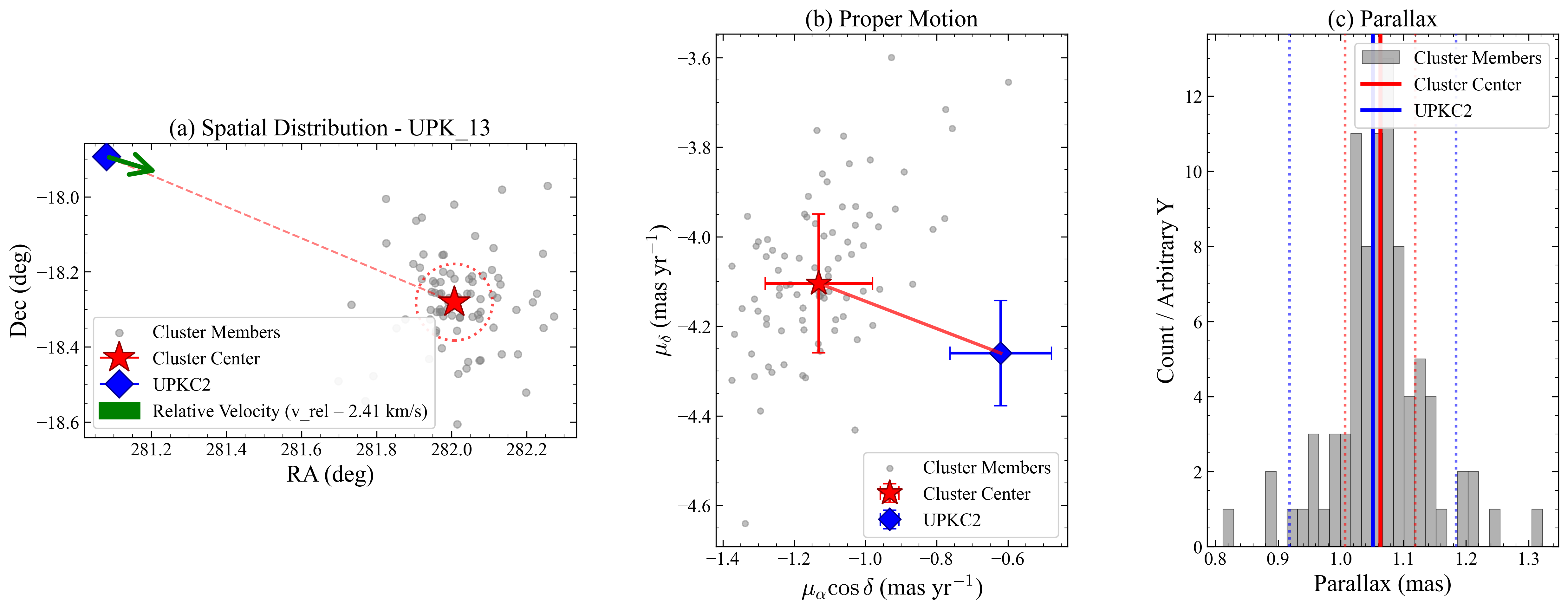}
\caption{Cluster membership diagnostics for UPK~13-c2.
(a)~Spatial distribution of UPK~13 members (grey circles) with the
cluster center (red star) and UPK~13-c2 (blue diamond).
(b)~Proper-motion diagram: UPK~13-c2 lies within $2.5\sigma$ of the
cluster mean proper motion.
(c)~Parallax histogram of cluster members; UPK~13-c2's adopted
parallax ($\pi_{\rm corr}=1.051\pm 0.133$~mas; catalog
value $1.013\pm0.132$~mas shown for reference) is within $0.2\sigma$ of the cluster
mean ($1.063\pm0.056$~mas), consistent with the cluster distance.}
\label{fig:cluster}
\end{figure*}

UPK~13-c2 satisfies three independent membership criteria for
UPK~13 (Figure~\ref{fig:cluster}): its proper motion lies within
$2.5\sigma$ of the cluster mean
($\chi^2_{\rm kin}=10.57$; Figure~\ref{fig:cluster}b), its
adopted parallax is within $0.2\sigma$ of the cluster mean distance
(Figure~\ref{fig:cluster}c), and the photometric WD$+$MS
classifier assigned it to the cluster catalog
\citep{Grondin2024}. The spatial offset of UPK~13-c2 from the
cluster center ($\chi^2_{\rm spat}=42.77$;
Figure~\ref{fig:cluster}a) is consistent with the known halo
extent of dynamically evolving open clusters.

We note that applying the \citet{MaizApellaniz2022}
zero-point correction shifts the target parallax \emph{toward}, not
away from, the cluster mean (the catalog $\pi=1.013$~mas becomes
$\pi_{\rm corr}=1.051$~mas, while the cluster centroid is at
$1.063$~mas). The membership case therefore strengthens, not weakens,
under the adopted parallax. We have also considered the possibility
of chance projection of a younger, unbound background binary along
this low-Galactic-latitude sightline: such a contaminant must
simultaneously match the cluster's proper motion within $2.5\sigma$
\emph{and} fall within the narrow adopted parallax range,
$\Delta\pi\lesssim 0.13$~mas. Using the UPK~13 field surface density
of Gaia DR3 sources at $G<18$ within $1\fdg5$ of the cluster centre
($\sim 6\times 10^3$~deg$^{-2}$) and a Galactic-disc proper-motion
dispersion of $\sim 7$~mas\,yr$^{-1}$ along the cluster vector at
$d\sim 1$~kpc, the joint coincidence probability is
$\lesssim 10^{-3}$ per cluster-sized field, sufficiently low that
field contamination is unlikely to be the dominant explanation.
Spectroscopic
confirmation of the systemic radial velocity and comparison
with the cluster mean $V_r\approx-7$~km\,s$^{-1}$
\citep{CantatGaudin2020} would provide the strongest direct membership
test.

%% ======================================================================
\section{Summary and Future Work} \label{sec:summary}

We have presented multi-band photometric evidence that UPK~13-c2,
classified as a high-confidence WD$+$MS
binary candidate, is more plausibly a late-K/early-M binary
(M1V$+$K9V, $M_{\rm tot}\approx 1.4\,M_\odot$ estimated from SED
bolometric luminosities; see Section~\ref{subsec:sed}) whose secondary is
periodically fully occulted by a misaligned circumbinary disk. Our main
findings are:

\begin{enumerate}
\item Square-wave eclipse. UPK~13-c2 shows a stable,
flat-bottomed eclipse ($P=36.71$~days, duration $\sim$10~days)
with an approximately achromatic $\sim$40\% flux decrement from
the optical through $W1$ ($0.47$--$3.4~\mu$m), with a reduced $W2$
depth ($\sim$28\%) attributable to disk thermal dilution, spanning
$\sim$6~years of observations.

\item Two-pronged evidence against WD$+$MS. The flat-bottomed,
multi-day-trapezoid morphology requires \emph{complete} occultation
of an extended ($R\!\sim\!0.7\,R_\odot$) component; a $\sim$9000-km
WD would cross the disk edge in $\lesssim$2~hours and cannot
naturally reproduce the observed 2.5-day ingress. Independently, the
decoupled-fit difference spectrum ($\chi^2_{\rm diff}\approx 0.9$
for K9V vs.\ $\chi^2_{\rm diff}\approx 45$ for a WD) strongly
disfavors the white dwarf hypothesis: the Rayleigh--Jeans
spectral morphology of a WD cannot naturally reproduce
the observed 44\% $W1$ and 28\% $W2$ eclipse depths.

\item Photometrically inferred MS$+$MS$+$disk solution. A
template-based late-K/early-M binary (M1V$+$K9V, estimated
$M_{\rm tot}\approx 1.4\,M_\odot$, subject to ${\sim}20\%$
systematic uncertainty from the adopted extinction,
parallax-based scaling, and broadband template matching) with a
cold ($\sim$220~K) circumbinary disk self-consistently reproduces
the eclipse morphology, approximately achromatic $\sim$40\% flux
decrement from optical through $W1$, $\sim$4000~K
difference-spectrum color temperature, and the
equilibrium dust temperature at the tidally truncated inner rim
($R_{\rm in}\approx 0.48$--$0.72$~AU).

\item Possibly the oldest known MS$+$MS disk-eclipsing binary candidate.
If cluster membership is confirmed, UPK~13-c2 may be
the oldest known disk-eclipsing binary in which both stars have
already arrived on the main sequence, by an order of magnitude
($\sim$316~Myr vs.\ $\lesssim$20~Myr for the previously known
MS$+$MS sample exemplified by Bernhard-2; \citealt{Hu2024}). This
would raise the question of how occulting circumbinary
material can persist to late main-sequence ages around
moderate-mass binaries.
\end{enumerate}

The MS$+$MS and WD$+$MS hypotheses make distinct, testable
predictions that can be addressed with modest observational effort.
High-resolution optical spectroscopy ($R\gtrsim 5000$) would help
determine whether the system is WD$+$MS or MS$+$MS while directly
yielding the dynamical masses and orbital eccentricity. These
fundamental kinematic parameters could test how the binary tidal
truncation radius relates to the inferred disk inner edge, and
whether circumbinary structures of this type can survive to
$\sim$300~Myr. Continued ground-based photometric monitoring of
the secular eclipse evolution may map the nodal precession of the
misaligned disk. Together, these follow-ups would help secure
the system's architecture and may establish whether UPK~13-c2 is
better described as a long-lived primordial reservoir or as a
massive collisional debris ring, providing a possible benchmark
for intermediate-age binary--disk interactions.

\begin{acknowledgments}
J.L. and C.L. are supported by the National Natural Science
Foundation of China (NSFC Grant No. 12033013).

We thank Zhecheng Hu (Tsinghua University) for his insightful comments and constructive suggestions that greatly improved this manuscript.
This work utilizes data from the Zwicky Transient Facility, the
Wide-field Infrared Survey Explorer, the ESA Gaia mission, NASA's
SPHEREx mission, Pan-STARRS1, and 2MASS.

Pan-STARRS1 (PS1) DR2 stack $i/z/y$ photometry was retrieved
from the PSPS catalog interface \citep{Chambers2016, Flewelling2020}.
SPHEREx Quick Release spectrophotometry was obtained from the
IPAC/IRSA SPHEREx Mission Archive.  Gaia DR3 data are from the ESA
Gaia Archive (\texttt{gea.esac.esa.int}); WISE / NEOWISE-R
single-exposure and unWISE coadded photometry are from the IPAC/IRSA
WISE archive; 2MASS data are from IPAC/IRSA; ZTF DR23 light curves
are from the ZTF Forced-Photometry Service at IPAC.  No MAST-hosted
data products were used in the final analysis.
\end{acknowledgments}

\vspace{5mm}
\facilities{ZTF, WISE, NEOWISE, Gaia, Pan-STARRS1, 2MASS, SPHEREx}

\software{astropy \citep{Astropy2013, Astropy2018, Astropy2022},
scikit-learn \citep{Pedregosa2011},
lightkurve \citep{Lightkurve2018},
emcee \citep{ForemanMackey2013},
corner \citep{ForemanMackey2016},
dustmaps \citep{Green2019}}

\bibliography{references}{}
\bibliographystyle{aasjournalv7}

\end{document}